\documentstyle[fleqn,12pt]{article}

\newcounter{abc}

\newcounter{Rom}

\newcounter{rom}

\begin{document}

\vspace*{3ex}

\Large
\begin{center}

   \vspace*{2ex}
     {\bf On the $K\bar{K}$ system as two-kaons system } \\

   \vspace*{2ex}
                   {\bf Jaroslav HRUBY}

   {\bf Institute of Physics AV CR,Czech Republic }
   {\bf e-mail: hruby.jaroslav@iol.cz  }
\vspace*{3ex}

\begin{abstract}
    The time evolution of the $K\bar{K}$ system as a two-qubit
    system is given.The effect which is interpreted as CP violation in neutral kaon decays is explained
     via violation of quantum correlations during time evolution of the $K\bar{K}$ system as
     two-kaons
    system and description is via Yang-Baxterization and unitary time dependent R-matrices to construct
    Hamiltonian, determining the time evolution of two-kaons system. The nonseparability ideas and criterion
     can be extended on all mixing state-antistate system and all CP violation cases in particle
     physics.

\end{abstract}
\end{center}

\vspace*{3ex}

\normalsize

\section{Introduction}

 We want to discuss the time evolution of the $K\bar{K}$ system as the a two-gubit system and show that experimental observation
of CP (or T) violation in particle physics can be explained as the
violation of correlation in state-antistate quantum system.
 The group formed by J.H.Christenson, J.W.Cronin, V.L.Fitch and
R.Turlay [1] found CP violation as an excess of events at the
pions $\pi^+\pi^-$ invariant mass corresponding to the $K_{L}$
state in the direction of the incoming beam and later the same
type of signal  was found by M.Holder et. al. [2] in the neutral
pion channel.

The results was interpreted as the effect of CP violation in
$K\bar{K}$ system and from this time CP and also T violation ( via
the validity of CPT theorem ), is accepted in elementary particle
physics (for review see [3]).

In this paper we want to show another possible explanation of these measurements
without the CP violation, using recent results from the quantum information theory [4].

At the beginning we want to say  that it is another story then
Einstein,Podolsky and Rosen problem which is in the decay of
$J^{PC}= 1^{--}$ vector meson into pair of entangled neutral K
mesons and was published in [5] and recently has been discussed in
many papers for example [6].

For our purpose we consider  the  $K\bar{K}$ as the two-kaons
system, what in quantum information is a two-qubit system realized
by the state $\mid K\rangle$ and $\mid \bar K\rangle $.

A system of such two-kaons is a four-dimensional Hilbert space
$H_4=H_2\otimes H_2$ having orthonormal basis

$\{\mid K\rangle \mid K\rangle, \mid K\rangle \mid \bar K\rangle,
\mid \bar K\rangle \mid  K\rangle, \mid \bar K\rangle \mid \bar
K\rangle \}$.

 We also write

$\mid K\rangle\mid K\rangle = \mid KK\rangle = KK , \mid K\rangle
\mid \bar K\rangle = \mid K \bar K\rangle = K\bar K$ etc.

    A state of two-kaons system is a unit-length vector
\begin{equation}
  {a}_0 |KK\rangle  + {a}_1 |\bar{K}K\rangle + {a}_2 |K\bar{K}\rangle  + {a}_3 |\bar{K}\bar{K}\rangle  ,
\end{equation}
so it is required ${|{a}_0|}^2 + {|{a}_1|}^2 + {|{a}_2|}^2 +
{|{a}_3|}^2  = 1$.

The construction of the Hamiltonian determining the time evolution
of such system can be obtained via Yang-Baxterization the unitary
R(x)-matrices in an eight-vertex model [4].
 In the braid group representation, where b
matrices have to satisfy braided relation :
\begin {equation}
 b_{i}b_{i+1} b_{i} = b_{i+1}b_{i}b_{i+1}, 1\leq i\leq
 n-1, b_ib_j = b_jb_i, \mid i-j\mid > 1.
\end{equation}
The quantum Yang-Baxter equation in R-matrices can be written as
follows:
\begin {equation}
 \Re_{i}(x)\Re_{i+1}(xy) Re_{i}(y)   = \Re_{i+1}(y)\Re_{i}(xy)\Re_{i+1}(x) ,
\end{equation}
with the asymptotic condition $\Re(x=0) = b$. The b-matrix and
R-matrix are $n^2 \times n^2$ matrices acting on $V \otimes V$,
where V is an n-dimensional vector space.

As b and $\Re$ acts on
the tensor product $V_{i} \otimes V_{i+1}$ we denote them $ b_{i}$
and $\Re_{i}$, respectively.

 The association of a unitary operator with a braid that respects
 the topological structure of the braid and allows exploration of
 the entanglement properties of the operator.The entanglement
 between two physical states or two-qubit states are known and
 play crucial role in quantum physics and QIS.

 We can see a state $\Psi\in H_4$
\begin{equation}
\Psi=\frac{1}{2}( |KK\rangle  + |K\bar K\rangle +  |\bar K
K\rangle + |\bar K\bar K\rangle )= \frac{1}{\sqrt{2}}(|K\rangle +
|\bar K\rangle)\frac{1}{\sqrt{2}}(|K\rangle + |\bar K\rangle) ,
\end{equation}
of a two-kaons system is decomposable, because it can be written
as a product of states in $H_2$. A state that is not decomposable
is entangled.

 Consider the unitary matrix
\begin{equation}
  \overline{R}=\left( \begin{array}{cccc}
                              a_0 & 0 & 0 & 0\\
                              0 & 0 & a_3 & 0\\
                              0 & a_2 & 0 & 0\\
                              0 & 0   & 0 & a_1   \end{array} \right) ,
\end{equation}
defines a unitary mapping, whose action on the $\Psi$ basis is
\begin{equation}
    \Psi=\overline{R}(\psi\otimes\psi) =       \overline{R} \left( \begin{array}{c}
                              |KK\rangle\\
                              |K\bar K\rangle\\
                              |\bar K K\rangle\\
                              |\bar K\bar K\rangle  \end{array}\right) = \left( \begin{array}{c}
                              a_0|KK\rangle\\
                              a_3|\bar K K \rangle\\
                              a_2|K \bar K\rangle\\
                              a_1|\bar K\bar K \rangle  \end{array}\right) .
\end{equation}

For $a_0a_1 \neq a_2a_3$ the state is entangled.For example the
state $\frac{1}{\sqrt{2}}(|\bar K K\rangle +|K\bar K\rangle)$ is entangled.

In physical experiments we have in every time two-kaons system.For
example we can see it in the regeneration of short-living K-meson
($K_S$). In a K–meson beam, after a few centimeters, only the
long–lived kaon ($K_L$) state survives. But suppose we place a
thin slab of matter into the beam then the short–lived state $K_S$
is regenerated because the K and $\bar K$ components of the beam
are scattered/absorbed differently in the matter.

Se in every moment we have $K\bar{K}$ system and we have to study
the time evolution of such system rather then system of one-kaons
states, as is usual:

ordinary it is the singlet short and long living one-kaon states
$K_S\equiv\mid S\rangle$ and $K_L\equiv\mid L\rangle$ so that
quantum mechanical states:
\begin{equation}
   \mid S\rangle = \frac{1}{\sqrt{2}}\{ \mid K\rangle  + \mid \bar K\rangle \} , \label{1}  \\
   \mid L\rangle = \frac{1}{\sqrt{2}}\{ \mid K\rangle  - \mid \bar K\rangle \} , \label{2}  \\
\end{equation}
where $ \mid S\rangle=S $ is state with strangeness $\hat{S}=+1$
and $CP=+1$ and $\mid L\rangle=L$ is state with strangeness
$\hat{S}=+1$ and $CP=-1$.

The time evolution of kaons is ordinary studied as

\begin{equation}
\mid K(t)\rangle = \frac{1}{\sqrt{2}}[U_S(t) \mid S(0)\rangle  +
U_L(t)\mid L(0)\rangle] ,   \\
\end{equation}
what is again a mixture of $K\bar{K}$ :
\begin{equation}
= \frac{1}{2}[U_S(t) + U_L(t)] \mid K(0)\rangle +
\frac{1}{2}[U_S(t) - U_L(t)]\mid\bar{K}(0)\rangle .
\end{equation}

The $\bar{K}$ evolves at proper time t into
\begin{equation}
\mid\bar{K}(t)\rangle = \frac{1}{\sqrt{2}}[ U_S(t) \mid S(0)\rangle - U_L(t)\mid L(0)\rangle ],   \\
\end{equation}
what is again a mixture of $K\bar{K}$:
\begin{equation}
= \frac{1}{2}[U_S(t) - U_L(t) ] \mid K(0)\rangle + \frac{1}{2}[U_S(t) + U_L(t)]\mid\bar{K}(0)\rangle,
\end{equation}

where $U_{S,L}(t)=\exp^{-{\alpha}_{S,L}}t$,
$\alpha_{S,L}=\frac{1}{2}\gamma_{S,L} + im_{S,L}$, where
$\gamma_{S,L}$ and $m_{S,L}$ denote the decay rate and mass,
respectively, of the S(L) kaon and units $\hbar=c=1$ have been
adopted.

The pair of two quantum mechanical one-states of an initial $CP
=\pm1$ kaon evolves at proper time t has form:
\begin{equation}
\mid S(t)\rangle = U_S(t) \mid S(0)\rangle=
U_S(t)\frac{1}{\sqrt{2}}\{ \mid K(0)\rangle  +
\mid \bar{K}(0)\rangle \} \ ,
\end{equation}
\begin{equation}
\mid L(t)\rangle = U_L(t) \mid L(0)\rangle=
U_L(t)\frac{1}{\sqrt{2}}\{ \mid K(0)\rangle  -
\mid \bar{K}(0)\rangle \} \ .  \\
\end{equation}

The K beam oscillates with frequency $\frac{m_L-m_S}{2\pi}$. The
oscillation is clearly visible at times of the order of a few
$\tau_S$, before all $K_S$ have died out leaving only the $K_L$ in
the beam. So in a beam which contains only K mesons at the time $t
= 0$ the $\bar K$ will appear far from the production source
through its presence in the $K_L$ meson with equal probability as
the K meson. A similar feature occurs when starting with a $\bar
K$ beam.So we can conclude that in reality we have two-kaons
quantum states, rather then one-quantum kaons states.

Now we show a time evolution of two-kaons system analogously as
two-qubit system.

\section{Time evolution of two-kaons system}

As is usually done we will assume that there is some unitary
operator $U_t$ mapping $H_4\rightarrow H_4$ which depends on the
time t, and describes the time evolution of the two-kaons system.
In other words, by denoting the state of two-kaons system
\begin{equation}
\Psi  = {(\mid KK, \mid K \bar K\rangle, \mid \bar K K\rangle,
\mid \bar K \bar K\rangle )}^T
\end{equation}
there is a unique self-adjoint operator H such that
\begin{equation}
U_t = e^{-itH}
\end{equation}
and the time evolution can be expressed as
\begin{equation}
\Psi (t)= U_t \Psi (0) = e^{-itH} \Psi(0).
\end{equation}

The Hamiltonian determining the time evolution of $\Psi$ is
constructed via Yang-Baxterization of the one solution of the
braid group representation (BGR) for the eight vertex model and
the corresponding unitary R-matrices ($\Re$) [4].

 In the quantum Yang-Baxter eq.(QYBE) the  solution of R-matrices usually depends on the
 deformation parameter q and the spectral parameter x.

 Taking the
 limit of $x\rightarrow0$ leads to the braided relation from
 the QYBE and the BGR b-matrices from the R-matrices.

The  Yang-
 Baxterization means to construct the $\Re(x)$ matrices from a given BGR
 b-matrices.
    For our purpose we use the BGR b-matrices of the eight-vertex model, which have the form
    [4]:
\begin{equation}
 b\pm   = \left( \begin{array}{cccc}
                              1 & 0 & 0 & q\\
                              0 & 1 & \pm1 & 0\\
                              0 & \mp1 & 1 & 1\\
                        -q^{-1} & 0    & 0 & 1  \end{array} \right)
                        .
\end{equation}

    It has two eigenvalues $\lambda_{1,2}= 1\pm i $. The
    corresponding $\Re (x)$-matrices via Yang-Baxterization is obtained
    to be
    \begin{equation}
 \Re_{\pm}(x) = b + x \lambda_1 \lambda_2 =
 \left( \begin{array}{cccc}
                              1+x & 0 & 0 & q(1-x)\\
                              0 & 1+x & \pm(1-x) & 0\\
                              0 & \mp(1-x) & 1+x & 1\\
                    -q^{-1}(1-x)& 0    & 0 & 1+x  \end{array}
                    \right).
\end{equation}

Assume the spectral parameter x and deformation parameter q to be
complex numbers, then the unitary condition for R-matrices leads
to the  equations (see [4]) which specify x real and q living at a
unit circle.

    Introducing the new variables $\theta,\varphi$ as follows
\begin{equation}
\cos\theta = \frac{1}{\sqrt{1+x^2}}, \sin \theta =
\frac{x}{\sqrt{1+x^2}}, q = e^{i\varphi}\label{1.1}
\end{equation}

the R-matrices has the form

\begin{equation}
\Re_\pm (\theta) = \cos(\theta)b_{\pm}(\varphi) +
\sin(\theta)b_{\pm}^{-1}(\varphi)\label{1.1}
\end{equation}

A method of constructing Hamiltonian from the unitary R-matrices
is known:
 the spectral parameter x play a role of time  for wave
function $\Psi$ specified by the R-matrices so we can write
\begin{equation}
\Psi(t)= \Re (t)\Psi
\end{equation}
 and the "time-dependent" Hamiltonian $H(t)$ has the form
\begin{equation}
H(t)= i\frac{\partial}{\partial t}
(\varrho^{-\frac{1}{2}}\Re)\varrho^{-\frac{1}{2}}{\Re}^{-1}(t),
\end{equation}
where $\varrho$ is
\begin{equation}
\varrho = \Re(t) \Re(t^{-1}) = (q^2 + q^{-2} - t - t^{-1} ),
\end{equation}
 and for
$\Psi(t)$ controlled by the unitary $\Re (t)$- matrix the
Schr\"{o}dinger equation corresponding to the time evolution is
valid
\begin{equation}
i\frac{\partial}{\partial t} \Psi (t)= H(t)\Psi(t).
\end{equation}

In this case of eight-vertex model we have the for
"time-dependent" Hamiltonians $H_{\pm} (t)$ the following
expression
\begin{equation}
H_{\pm}(t)= i
(\varrho^{-\frac{1}{2}}\Re)\varrho^{-\frac{1}{2}}{\Re}^{-1}(t)=
-\frac{i}{1+t^2}{b_{\pm}}^2.
\end{equation}

This Hamiltonians are invariant under discrete time transformation
$t\rightarrow -t$.

The time-independent Hamiltonian $H_\pm$ in $t=1$ has the form
\begin{equation}
 H_\pm =  -\frac{i}{2}{b_\pm}^2(\varphi) = \frac{i}{2} \left( \begin{array}{cccc}
                              0 & 0 & 0 & -e^{i\varphi}\\
                              0 & 0 & \mp1 & 0\\
                              0 & \pm1 & 0 & 0\\
                        e^{-i\phi} & 0    & 0 & 0  \end{array} \right)
                        .
\end{equation}

So we get
\begin{equation}
H_\pm \Psi(t) = {(-e^{-i\varphi}\bar K\bar K,\mp \bar K K,\pm
K\bar K,e^{-i\varphi}KK )}
\end{equation}

It means that during time evolution of two-kaons system we
obtained some deformation if $\varphi\neq0$ of two kaon-states KK
and $\bar K \bar K$ which change maximum entanglement in Bell
states $\Phi$.

\section{Violation of correlations between kaon states rather then CP violation}
The action of the braid group representation $b_\pm(\varphi)$ on
the two-kaons state $\Psi$ yield the Bell states with the phase
factor $e^{i\varphi}$

\begin{equation}
 b_\pm(\varphi) \left( \begin{array}{c}
                              |KK\rangle\\
                              |K\bar K\rangle\\
                              |\bar K K\rangle\\
                              |\bar K\bar K\rangle  \end{array}\right) =\frac{1}{\sqrt{2}} \left( \begin{array}{c}
                              |KK\rangle + e^{i\varphi}|\bar K\bar K\rangle\\
                              |\bar K K\rangle \pm |K \bar K\rangle\\
                              \mp|K\bar K\rangle + |\bar K K\rangle\\
                              -e^{-i\varphi}|KK\rangle + |\bar K\bar K \rangle \end{array}\right) .
\end{equation}

which shows that $\varphi=0$ leads to the Bell states, the maximum
of entangled states:

\begin{equation}
\Phi_{1,2}=\frac{1}{\sqrt{2}}(|KK\rangle \pm \bar K\bar K\rangle),
\end{equation}
\begin{equation}
\Phi_{3,4}=\frac{1}{\sqrt{2}}(|\bar K K\rangle \pm |K\bar
K\rangle).
\end{equation}

These states are different then the states $K_{1...4}$ which were
introduced in [5].There the states $K_1=K_S$ and $K_2=\bar{K_S}$
are identical.

The necessary codefinition of strangeness $\hat{S}$ and CP can be
rigorously justified by applying local realism to kaons belonging
to a correlated kaon pairs in Bell states.

If we have the action of $\hat{S}$ and CP operators the following:
\begin{equation}
\hat S K= + K  , \hat S\bar K = - \bar K ,
\end{equation}
and
\begin{equation}
(CP) K= - \bar K  , (CP)\bar K = - K,
\end{equation}

then for the four  states $\Phi_{1,...4}$ we have:

\begin{equation}
S\Phi_{1,2}= + \Phi_{1,2}  , CP\Phi_{1,2}= \pm \Phi_{1,2}
\end{equation}

\begin{equation}
S\Phi_{3,4}= - \Phi_{1,2}  , CP\Phi_{3,4}= \pm \Phi_{3,4}.
\end{equation}

So we have two pairs with $CP=+1$ and $CP=-1$

 -the  pair with $\hat{S}\pm1,CP=+1$: $\Phi_1, \Phi_3$,

-the second pair with $\hat{S}\pm1,CP=-1$: $\Phi_2,\Phi_4$.

Local realism allows one to attribute elements of reality to each
one of two kaons belonging to the Bell states $\Phi_{1...4}$.
While dichotomic element of reality $CP=\pm1$ describes the time
independent property, the dichotomic element of reality $S=\pm1$
describes an instantaneous property.The necessary codefinition of
S and CP can be rigorously justified by applying local realism to
a correlated kaon pairs in the entangled states $\Phi_{1...4}$.
The experimental measurement on the entangled state for example on
 $\Phi_1=\frac{1}{\sqrt{2}}(|KK\rangle + \bar K\bar K\rangle)$
will give K or $\bar K$ with a probability one half. Entanglement
is here in the time. K moves in t and $\bar K$ like K in -t .

The deformation parameter $\varphi$ appears only for $\Phi_1$ and
$\Phi_4$ and gives the deformation of the correlation it means
that states are not the maximum entangled states like for case of
Bell states $\varphi=0$.

So during the time evolution of the two-kaons system the time
symmetry (and also CP) is preserved and correlations in
entanglement states are deformed so that there is not the maximum
entanglement [4].

\section{Conclusions}

The concept of entanglement for pure quantum states was
established in the early days of quantum mechanics, but the use of
the concept for mixed states as two-kaons analogously like
two-qubits is relatively recent [4].

Here we show the new application of these results, namely
explanation the effect of CP violation in K$\bf\bar{K}$ system via
description of two-kaons system analogously two-qubit systems in
quantum information science . Our result is that there can be no
violation of CP and T in QM.

Physically it is more natural to explain the experimental
observations which are coming from the mixtures in
states-antistates like deformation of the correlation or
entanglement then the violation of operators of discrete
relativistic invariance.

The nonseparability ideas and criterion
     can be extended on all mixing and CP violation cases in particle
     physics.
     and experimental results can be explained without the CP violation.
In all cases there exist a correlated pair $K\bar{K},B\bar{B},
...$ and effects of CP violation can be explained via QM
nonseparability. It can be applied also on qubit-antiqubit states
[7] where antiqubit states can be understood as qubit states
moving information backward in time. This deformation of
entanglement in particle-antiparticle system can explain also  why
antimatter disappeared in the beginning of the Universe.

On the end we want to say that the more natural statement is to
believe that in QM is no CP violation and there is only QM
non-separability of the wave function and deformation of the
maximum entanglement states what is experimentally verified.

\end{document}